Review

# Digital Health Innovations for Screening and Mitigating Mental Health Impacts of Adverse Childhood Experiences: Narrative Review

Brianna M White[1], MPH; Rameshwari Prasad[1], MBBS, MPH; Nariman Ammar[2], MSc, PhD; Jason A Yaun[3], MD; Arash Shaban-Nejad[1], MSc, MPH, PhD

[1]Department of Pediatrics, Center for Biomedical Informatics, The University of Tennessee Health Science Center-Oak Ridge National Laboratory, Memphis, TN, United States
[2]School of Information Technology, Illinois State University, Normal, IL, United States
[3]Methodist Le Bonheur Healthcare, Memphis, TN, United States

**Corresponding Author:**
Arash Shaban-Nejad, MSc, MPH, PhD
Department of Pediatrics, Center for Biomedical Informatics
The University of Tennessee Health Science Center-Oak Ridge National Laboratory
50 N Dunlap St
Memphis, TN, 38103
United States
Phone: 1 9012875836
Email: ashabann@uthsc.edu

## Abstract

**Background:** Exposures to both negative and positive experiences in childhood have proven to influence cardiovascular, immune, metabolic, and neurologic function throughout an individual's life. As such, adverse childhood experiences (ACEs) could have severe consequences on health and well-being into adulthood.

**Objective:** This study presents a narrative review of the use of digital health technologies (DHTs) and artificial intelligence to screen and mitigate risks and mental health consequences associated with ACEs among children and youth.

**Methods:** Several databases were searched for studies published from August 2017 to August 2022. Selected studies (1) explored the relationship between digital health interventions and mitigation of negative health outcomes associated with mental health in childhood and adolescence and (2) examined prevention of ACE occurrence associated with mental illness in childhood and adolescence. A total of 18 search papers were selected, according to our inclusion and exclusion criteria, to evaluate and identify means by which existing digital solutions may be useful in mitigating the mental health consequences associated with the occurrence of ACEs in childhood and adolescence and preventing ACE occurrence due to mental health consequences. We also highlighted a few knowledge gaps or barriers to DHT implementation and usability.

**Results:** Findings from the search suggest that the incorporation of DHTs, if implemented successfully, has the potential to improve the quality of related care provisions for the management of mental health consequences of adverse or traumatic events in childhood, including posttraumatic stress disorder, suicidal behavior or ideation, anxiety or depression, and attention-deficit/hyperactivity disorder.

**Conclusions:** The use of DHTs, machine learning tools, natural learning processing, and artificial intelligence can positively help in mitigating ACEs and associated risk factors. Under proper legal regulations, security, privacy, and confidentiality assurances, digital technologies could also assist in promoting positive childhood experiences in children and young adults, bolstering resilience, and providing reliable public health resources to serve populations in need.

## Introduction

Exposures to both negative and positive experiences in childhood have proven to influence cardiovascular, immune, metabolic, and neurologic function throughout an individual's life [1,2]. As such, adverse childhood experiences (ACEs) could have severe consequences on health and well-being into adulthood [3,4]. ACEs are defined as traumatic childhood events, including abuse (physical, emotional, or sexual), neglect (physical or emotional), household dysfunction, caregiver or parental loss, and experiences of trauma around other social determinants of health [5]. Repeated exposure to a broad variety of traumas in early childhood places particular strain on mental health and is correlated with illicit substance use disorders, depression, obesity, and increases in suicidal ideation into adulthood [6]. Due to recurrent stress response activation before the age of 18 years, children and young people with a history of multiple ACEs are at higher risk of developing neuropsychiatric and emotional disorders, including posttraumatic stress disorder (PTSD), anxiety or depression, attention-deficit/hyperactivity disorder (ADHD), and schizophrenia [7]. However, the relationship between ACEs and behavioral problems is sometimes characterized as bidirectional, with a disorder diagnosis preceding the incidence of ACEs [8]. Regardless of the bidirectional potential, estimates show that more than 50% of children and young people in the United States have reported experiencing at least 1 ACE before their 18th birthday, emphasizing a growing need for intervention, including the continued exploration of the role of positive childhood experiences (PCEs) [9].

Pressing the issue, the global COVID-19 pandemic influenced a marked rise in reported ACEs. A recent Adolescent Behaviors and Experiences Survey conducted by the Centers for Disease Control and Prevention found that 73% of high school–aged students reported experiencing at least 1 adverse event during the pandemic period [10]. The study found that exposed students had a significantly higher prevalence of mental health complaints and an increase in suicidal ideation as compared to adolescents reporting no ACEs [10]. In the wake of this escalation, a national emergency in child and adolescent mental health was declared by the American Academy of Child and Adolescent Psychiatry, the American Academy of Pediatrics, and the Children's Hospital Association in late 2021 [11]. The advisory sheds light on the severe mental health consequences of the pandemic, noting the extraordinary weight of children and young people being forced to socially isolate and distance themselves from peers, cope with unprecedented grief, and adapt to digital learning environments [11]. Additionally, more than 140,000 children and young people were reported to have experienced the death of a parent or caregiver due to COVID-19 infection in the first year of the pandemic alone [12]. While associated concerns preceded the COVID-19 pandemic, recent exacerbation highlights the occurrence and related effects of adverse events on the mental health of children and young people as a critical public health issue.

Recent research efforts have focused on understanding how PCEs may help to counter the risks of ACEs [6,13]. Emerging evidence suggests that early access to safe, stable, nurturing environments; positive attachments with caregivers, peers, and teachers; interpersonal connection; opportunities for emotional growth; school engagement; and connection to culture and community are among PCEs linked to strengthening childhood resilience and overall health and well-being across the life span [14,15]. Such experiences aid in the vital development of a sense of belonging and connectedness, which often serve as a helpful buffer to the negative effects of ACEs [16].

As digital use by children and adolescents has exponentially increased over the past decade [17], the potential for digital health interventions is promising. The application of digital health technologies (DHTs) has the potential to assist in screening and mitigation of the profound impacts of early life experiences and improve the well-being of children and young people [18]. DHTs have been emerging as transformative solutions to challenges in health care delivery since the early 2000s and are considered to promote equitable, affordable, and universal access to patients and health care providers [19]. The technologies, such as mobile health (mHealth), health information technologies, wearable smart devices, wireless medical devices, personalized medicine, and telemedicine platforms, offer innovative tools to enhance population health and public health responses [20].

When used to mitigate the profound and long-lasting effects of adverse experiences on children and young people, comprehensive, trauma- and evidence-informed digital tools can increase treatment accessibility, strengthen prevention and screening resources, and bolster available support for individuals most vulnerable to ACEs. Given the ever-increasing ubiquity of digital tools, the distribution of DHTs has the potential to reach children and young people and their families on a more intimate level. The innovative development of DHTs for ACE intervention could open new therapeutic pathways for children and adolescents, and such DHTs could also serve as knowledge reservoirs for parents and caregivers to aid in promoting positive experiences in childhood, bolstering resilience [21]. As research examining the role of PCEs continues to grow, increased development of DHTs could provide new opportunities for novel approaches to trauma-informed care in both the clinical setting and in the home. With the help of resources and recommendations on demand, families could readily access personalized messages in the form of texts and multimedia aimed at countering the negative health effects associated with ACEs outside of the traditional clinical setting. This could alleviate several barriers to care, including financial constraints (ie, insurance and copay challenges), geographical or physical inaccessibility (ie, health care access in rural settings, transportation), additional childcare for large families, or inadequate health care resources (ie, provider shortages), providing immediate treatment from a distance [22].

Ultimately, promoting the use of DHTs expands care access and strengthens patient and family engagement. Accordingly, the objective of this literature review was to evaluate and identify means by which existing digital solutions may be useful in mitigating the mental health risk factors and consequences associated with the occurrence of adverse experiences in childhood and adolescence and preventing ACE occurrence, including by highlighting knowledge gaps or barriers for DHT implementation for vulnerable populations. We argue that DHTs could help to prevent and mitigate the lasting effects of ACEs on children and young people, bolstering resilience, promoting PCEs, and alleviating the burden of ACE-associated impacts on health and well-being into adulthood.

# Methods

## Search Strategy

We performed a search of existing literature, incorporating peer-reviewed studies that were identified from 3 databases (PubMed, SSRN, and Google Scholar) and published between August 2017 and August 2022, to allow for evaluation of the most recent applications of DHTs. Our search terms were used based on the thesaurus and keywords, including "digital health," "adverse childhood experiences," "artificial intelligence," "domestic violence," "abuse," "mental health," "family environment," "COVID-19," "suicide," "depression," "anxiety," "social determinants of health," "personal health library," "precision medicine," "substance abuse," "alcohol use," and "post-traumatic stress disorder." Terms and keywords were combined to comprise search phrases, allowing for a more sensitive search. Search phrases included "the application of AI and DH to prevent ACEs," "role of DH and AI for mental health," "knowledge gap in digital health," "significance of DH in mental health," "specific AI solutions for ACE risks," "DH interventions in adolescence," "children who need special care," "machine learning application," and "electronic health record." Additionally, reference lists from selected papers were searched for possible matches and inclusion.

## Inclusion and Exclusion Criteria

Studies were included for review if they met the following eligibility criteria: (1) published from August 2017 to August 2022, (2) explored the relationship between digital health interventions and mitigation of negative health outcomes associated with mental health in childhood and adolescence, (3) examined prevention of ACE occurrence associated with mental illness in childhood and adolescence, and (4) conducted in the United States. Studies were excluded if they were not original research or were not available in the English language.

## Study Selection, Data Extraction, and Analysis

The initial electronic database search generated papers whose titles or abstracts were collectively screened by 2 reviewers (RP and BMW) to ensure fit to the specified eligibility criteria. The following data were extracted: title, lead author, publication year, study purpose, study design, ACE risk factor indicated, and author's conclusion. Extracted data were collated and stored on a Microsoft Excel spreadsheet (Microsoft Corp) coding matrix.

# Results

## Overview

The search of PubMed, SSRN, and Google Scholar and reference lists of each selected paper yielded a total of 105 results. In total, 87 studies were eliminated because they did not meet the inclusion criteria. After title or abstract screening and full-text review, a total of 18 studies were selected for data extraction and analysis, as demonstrated in Figure 1.

The oldest paper included in this study was published in August 2017, with others as recent as May 2022. Study characteristics and general information on papers selected for review are included and outlined in Multimedia Appendix 1. A total of 100% (18/18) of included papers positively reported the use of digital solutions for ACEs.

**Figure 1.** Paper selection process.

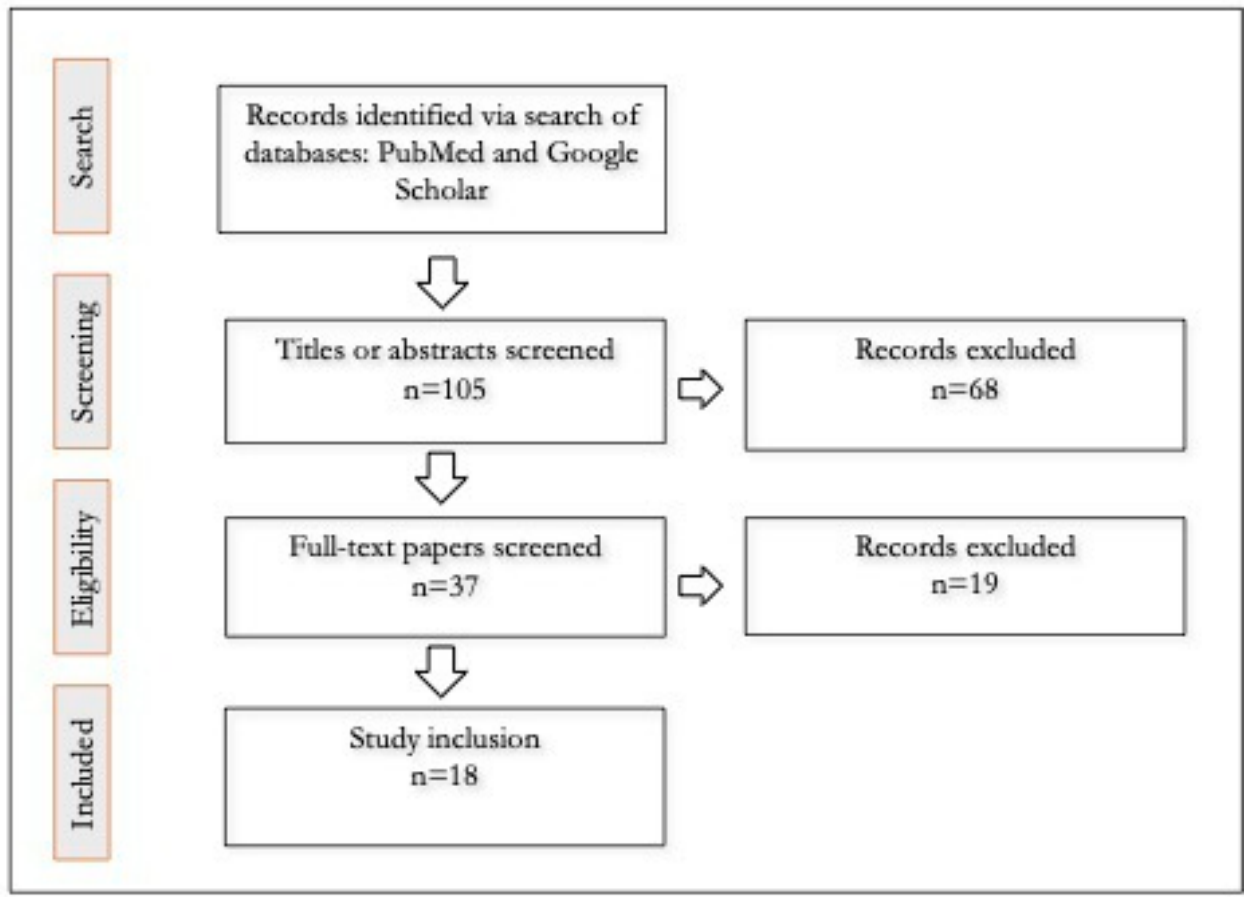

## Use of DHTs for the Management of ACE-Related Mental Health Outcomes

Overall, 44% (8/18) of our examined papers reported the use of DHTs to identify indicators of mental illness and reduce the negative effects of such adverse experiences. Studies assessed the suitability, usability, and acceptability of digital health interventions for the management of mental health illness with findings in support of the effectiveness of computerized cognitive behavioral therapies [23-28]. Such tools are argued to help diagnose, predict, and prevent negative outcomes of mental health illness due to ACEs. Examined studies argue that digital solutions, such as educational mobile apps, directed at children and young people have the potential to be important assessment, management, and treatment tools to provide efficient access to health services and promote self-help in hours of need [23,25,28]. Similarly, another study found that children and young people responded well to structured digital peer-to-peer interactions and strength-based mHealth coaching [26]. Moreover, collective findings highlight the potential for the use of social media–based digital intervention for children and young people at risk of negative mental health outcomes associated with ACEs [27,28].

## Posttraumatic Stress Disorder

Children exposed to traumatic ACEs are reported to have a higher incidence of PTSD compared to their peers without repeated stress [29,30]. One study from our search found promise in the use of artificial intelligence and longitudinal data to identify and treat children with PTSD as a result of childhood traumas [30]. Lekkas and Jacobson [30] found the use of geometric Global Positioning System data to measure daily minutes spent and the radius traveled away from home was an effective way to detect "avoidance symptoms" displayed by high-risk cohorts with a history of abuse and neglect. The study found that, without clinical predictors, the deployed passive monitoring tools successfully diagnosed PTSD more than 80% of the time [30].

## Suicidal Behavior or Ideation

From our search, 22% (4/18) of selected studies characterized opportunities for DHTs, machine learning, and electronic health record restructuring to predict the suicidal risk in children and young people having experienced 1 or more ACEs.

Overwhelmingly, suicide rates are 30 times higher among adults with more than 4 ACEs, highlighting an urgent need for innovative solutions [31]. Findings from examined papers suggest that DHTs could play key roles in identifying predictors of suicidal behavior, violence, and unintentional injury. Chen and Chan [32] found that digital health interventions have great potential to reduce unintentional injury, violence, and suicide in children and young people who have witnessed or experienced 1 or more traumatic events. The study argues that future work should focus on intervention to increase access to comprehensive digital technologies for those who have experienced or witnessed abusive behaviors in childhood or adolescence [32]. Notably, conclusions from another investigation suggest that predictive models of suicide risk of children and adolescents, by using demographics, comorbidity diagnosis codes, laboratory test results, and medications from clinical records models, show good performances for estimation of short-term and long-term risks and identify significant predictors, which may assist in clinical practices [33].

Others found that DHTs can be used to successfully predict, and potentially prevent, suicide risk among adolescents [33-35]. One used machine learning to successfully detect 53%-62% of suicide-positive behavior with 90% specificity [34]. Findings support growing evidence that the innovative use of electronic health records can be useful in developing reliable predictive models for suicide risk among children and young people [34]. Similar findings were expressed by Walsh et al [35], whose findings further suggested that routinely collected clinical data can be leveraged to predict suicide risk in children and young people with a history of ACE with computational algorithms. However, it is important to note that these findings are not always generalizable on a global scale, as only a small proportion of existing digital platforms in low- and middle-income countries are evidence-based and frequently demonstrate low effectiveness due to factors such as societal mistrust [24]. As such, widespread adoption and scale-up of digital mental health interventions will require more rigor in research and study before widespread global implementation.

### Anxiety or Depression

Anxiety is characterized by the reflection of internal feelings of discomfort expressed with uneasiness, and prolongation of these feelings can cause severe depression in children and young people [36]. Unfortunately, there are only a small number of studies available examining the correlation between ACEs and anxiety, with far fewer deciphering noticeable differences between anxiety and depression for children and young people having experienced 4 or more ACEs [36]. However, of the results from our search, 16% (3/18) indicated possible uses of digital health to support and care for children experiencing anxiety. Findings from one examined study suggest that digital interventions aimed at the improvement of anxiety symptoms in children and young people can be transformative in supplementing traditional treatment plans [37]. Khanna and Carper [37] found that digital mental health interventions, such as web- or cloud-based therapeutic programs, mobile apps, and virtual reality simulations, are reliable mitigation tools and should be used to supplement and maximize anxiety treatment completion.

This sentiment is echoed by both Williams and Fried et al [38,39], who also found a significant opportunity for the employment of digital interventions. Both argue that the development of mental health monitoring applications for anxiety and depression could allow for real-time recording of mental health symptoms, including perceived feelings and behaviors [38,39]. Conclusions suggest that population-wide use of such monitoring applications could reshape the clinical understanding and progression of mental health conditions, such as anxiety or depression, creating opportunities for targeted mitigation response [38,39].

### Attention-Deficit/Hyperactivity Disorder

Children and young people experiencing the effects of ACEs are at heightened risk of developing behavioral disorders such as ADHD [40]. From our search, 16% (3/18) of examined studies highlight the use of digital therapeutic interventions for the improvement of ADHD. Studies found that game-based digital therapeutic devices, such as the Food and Drug Administration–approved EndeavorRx, support positive outcomes when used to treat ADHD among children aged 8 to 12 years without the need for pharmacological intervention [41]. Kollins et al [42] found that daily game-like interaction with a digital therapeutic program (AKL-T01) improved inattention in children and young people diagnosed with ADHD. Similarly, Fried et al [39] found that the use of a digital mindfulness application showed a marked reduction in inattentiveness and anxiety for children aged 6 to 12 years with a history of ADHD. Additionally, results from a survey administered during the 4-week pilot study found a correlation between sleep patterns and both stress levels and attention span [39]. These findings could help to inform appropriate management plans and highlight the potential for the continual development of game-like digital interventions designed to exercise memory training and neurofeedback [41,42]. These innovative applications not only provide clinicians with attention assessments by measuring key stimuli and response variabilities but also supply users with continual intervention by sharpening perseverance and self-efficacy skills [42].

## Discussion

### Principal Findings

Immense challenges exist in overcoming ACEs in children and young people. Physical, sexual, or emotional abuse in childhood can severely impact the health of children and young people. Children who have experienced or witnessed more than 3 related types of abuse are at high risk for self-destructive behaviors, including substance abuse, suicidal ideation, and overall poor resilience [43].

As screening can be time-consuming and difficult to implement on a broad-reaching scale, DHTs may provide a scalable approach to capturing vulnerability and resilience factors in real time. We argue findings from our review may suggest that incorporating DHTs for managing the mental health consequences of repeated exposure to childhood trauma shows promise for building childhood resilience.

### Innovation

While arguably still in the early decades of meaningful implementation, there is pronounced potential for innovative digital technologies and interventions to identify indicators of mental illness and reduce the risk factors and negative effects of related adverse experiences. Mental illnesses related to ACEs are most commonly the result of enduring verbal, sexual, or physical abuse; neglect; parental separation or incarceration; familial substance abuse; domestic violence; poverty; peer rejection; and death or traumatic loss before age 18 years [44]. Additionally, many social determinants contribute to unfavorable familial conditions negatively affecting the health of children and young people, such as unemployment, food insecurity, single parenting, and unsafe neighborhood environments [44]. Support opportunities provided by DHTs supply health care providers with necessary mitigation tools for the extraction, processing, assessment, and analysis of patient-provided information for use in rapid response strategies [45-47]. Additionally, advancements in digital surveillance tools tasked with symptom-based geolocation could help pinpoint vulnerable communities for tailored mitigation, including comprehensive educational programs focused on the prevention of ACEs.

Perhaps harnessing the greatest potential are the innovation and scaling up of game-based therapeutic applications. As demonstrated in our results, these digital therapeutics have been shown to positively support the management and improvement of multiple mental health conditions (ie, PTSD, suicidal ideation, anxiety, depression, and ADHD) by engaging children and young people with unconventional treatment approaches [48]. For example, digital meditation and relaxation applications are therapeutic for children and young people with anxiety, depression, and sleeping problems, alleviating burdens related to increased stress factors and lack of adequate rest [39]. These technologies could also be used to facilitate resilience by promoting digital literacy and positive health information–seeking behavior and delivering precision health education or promotion. Digital interventions hold the potential to provide targeted resources to elevate PCEs, transforming early childhood experiences for those at the highest risk of facing adversity. DHTs may serve as a buffer against the effects of ACEs by broadening opportunities for exposure to positive experiences such as play, recognition, acceptance, praise, and other protective factors. Moreover, such digital technologies could prove to be advantageous in bolstering appointment and follow-up arrangements, patient screening and monitoring, and improvement of training and education for providers and decision makers. This bidirectional expansion could help bridge the gap between patient needs and care services.

## Limitations

While this review highlights the strengths and successes of digital solutions for mitigating ACEs, it also helps uncover the need for further exploration of their actionability. Challenges related to addressing gaps in digital health literacy, high-speed internet access, confidentiality, reliability, and trust remain. Ethics and practice guidelines are needed, and questions remain regarding the level and quality of therapeutic involvement needed to maximize treatment to bolster resiliency and positive outcomes in youth. Moreover, underreporting of ACEs and subsequent lack of treatment of associated consequences remain barriers due to the sensitive, deeply personal nature of mental health illness [49]. The growing body of evidence supports and encourages the development of personalized patient-provider communications, the fostering of personal relationships, and the counteraction of misinformation as solutions to decrease uncertainty and bolster patient-provider trust [50-52]. Likewise, there is room for comprehensive DHTs tasked with the creation and amplification of tailored messaging to target populations who display hesitancy to openly discuss mental health concerns with care providers. These tools could offer comprehensive educational material during care visits, complementing the therapeutic relationship of interpersonal communication and treatment and helping to bridge the gap between patient and provider when discussing mental health.

Furthermore, training and ACE awareness or educational campaigns to bolster proficiency for health care professionals could be achieved through structured digital learning and mHealth technologies. With proper training, health care professionals would be adequately equipped to both assess and accurately report ACE-related findings and effectively communicate prevention and response strategies to patients at risk. Although DHTs harness the potential for the screening and prevention of ACEs, it is also imperative to facilitate equitable access to these digital tools for vulnerable, harder-to-reach communities. Special attention should be given to developing DHTs with the capability to navigate and transcend many social determinants of health. As previously mentioned, challenges related to high-speed internet access may prevent families from fully harnessing the potential of DHTs. With this in mind, a standard of DHT development should include the ability to access tools and resources offline. Moreover, contextual information should be made available at every literacy level to account for a range of understanding. Overall, vulnerable populations must be considered during policy development and decision-making before implementing any architecture.

There is also a growing need for transparency, explainability [53], and understanding of digital health infrastructure for children and young people, particularly for those with special needs. Technologies aimed at children and young people should be standardized, so that messaging is consistent across varying platforms and information systems, as standardization could aid in mitigating confusion among the public when navigating mental health illnesses and other health disparities. Adequate funding is needed for institutions of learning (ie, grade schools and universities) to offer courses on digital health literacy, health information–seeking behaviors, ACE risk factors or consequences, big data analytics, and artificial intelligence. Additionally, there is room for participation and infrastructure development from private sector stakeholders (eg, information and communication technology companies) to facilitate efficient operationalization and optimization of digital health initiatives. The progressive growth of DHTs, especially in preventing mental health issues, is warranted to make solutions more accessible to a broader range of those in need.

## Conclusions

Successful development and deployment of digital solutions could reduce the consequences to mental health related to adverse childhood events; bolster childhood resilience; and influence future health policy, decision-making, and program implementation for ACE prevention and control [46,54].

From our findings, digital solutions, such as game-like mindfulness applications, have notably aided in the identification or treatment of PTSD, suicidal ideation, anxiety, depression, and ADHD in children at risk of or experiencing negative health effects associated with ACEs. While digital solutions should not be considered as a replacement for physical evaluation or treatment, the innovation of such tools could aid in identifying risk factors and preventing future complications and long-term illness or medical care dependence as a result of these conditions. We argue there is room for governmental and private sector stakeholder support of expansive, digital, open-source, modular DHTs with user-friendly graphical interfaces. Additionally, policy and data governance investments could support the tailored delivery of therapies and education efforts to ensure resource-limited populations have full access to these comprehensive technologies, preventing more severe negative outcomes.

Optimizing DHTs with the proper use of these technologies under regulatory control, screening, and prevention of the risks of ACEs can bolster public health response to mental health care for children and young people. Furthermore, consideration of the probable bidirectionality in the relationship between ACEs and mental health consequences when developing comprehensive digital tools could further help prevent associated impacts on well-being in adulthood, no matter the etiology of adversity. Regardless of origin or onset, prioritization of digital interventions targeting all underlying pathways could help to interrupt cycles of adversity and promote resilience for children experiencing mental illness. Future expansion of DHTs and their use could help bridge the gap between patients, providers, and researchers, increasing the number of PCEs and reducing associated morbidities and mortalities by enhancing screening and prevention of ACEs at earlier stages. Expanding and adopting such innovative, cost-effective digital solutions could have meaningful impacts on health care services for those affected by ACEs.

### Authors' Contributions

BMW and RP conceptualized the study and drafted, reviewed, and edited the manuscript. NA and JAY reviewed and edited the manuscript. AS-N drafted, reviewed, and edited the manuscript; supervised the study; and acquired funding.

### Conflicts of Interest

None declared.

### Multimedia Appendix 1

Summary table representing the characteristics of selected studies.
[DOCX File (Microsoft Word File), 22 KB-Multimedia Appendix 1]

## Abbreviations

**ACE:** adverse childhood experience
**ADHD:** attention-deficit/hyperactivity disorder
**DHT:** digital health technology
**mHealth:** mobile health
**PCE:** positive childhood experience
**PTSD:** posttraumatic stress disorder